\documentclass[twocolumn,amsmath,amssymb,superscriptaddress,longbibliography,reprint]{revtex4-1}

\newcommand{\IMSS}{Muon Science Laboratory, Institute of Materials Structure Science, High Energy Accelerator Research Organization (KEK-IMSS), Tsukuba, Ibaraki 305-0801, Japan}

\newcommand{\Sokendai}{Graduate Institute for Advanced Studies, SOKENDAI}
\newcommand{\MDXES}{MDX Research Center for Element Strategy, Institute of Integrated Research, Institute of Science Tokyo, Yokohama, Kanagawa 226-8501, Japan}
\newcommand{\MCES}{Materials Research Center for Element Strategy, Tokyo Institute of Technology, Yokohama, Kanagawa 226-8503, Japan}
\newcommand{\NIMS}{National Institute for Materials Science, Tsukuba, Ibaraki 305-0044, Japan}

\usepackage{multirow}
\usepackage{graphicx}
\usepackage[dvipdfmx]{color}
\usepackage{dcolumn}
\usepackage[version=3]{mhchem}
\usepackage{txfonts}
\usepackage{bm}
\usepackage{ulem}

\usepackage[utf8]{inputenc}
\usepackage[T1]{fontenc}
\usepackage{mathptmx}
\usepackage{etoolbox}
\usepackage[hypertex,colorlinks=true,linkcolor=black,citecolor=blue,filecolor=blue,urlcolor=blue,setpagesize=false,nesting=true]{hyperref}

\makeatletter
\def\@email#1#2{%
 \endgroup
 \patchcmd{\titleblock@produce}
  {\frontmatter@RRAPformat}
  {\frontmatter@RRAPformat{\produce@RRAP{*#1\href{mailto:#2}{#2}}}\frontmatter@RRAPformat}
  {}{}
}%
\makeatother

\begin{document}
\title{Relationship between local hydride ion dynamics and ionic conductivity in LaH$_{3-2x}$O$_x$\\  inferred from muon study}

\author{M.~Hiraishi}\affiliation{\IMSS}
\author{S.~Takeshita}\affiliation{\IMSS}\affiliation{\Sokendai}
\author{H.~Okabe}\affiliation{\IMSS}
\author{K. M. Kojima}\thanks{Present Address: Center for Molecular and Materials Science, TRIUMF, Vancouver, B.C., V6T2A3, Canada}\affiliation{\IMSS}
\author{A.~Koda}\affiliation{\IMSS}\affiliation{\Sokendai}
\author{S. Iimura}
\affiliation{\NIMS}
\author{K. Fukui}\thanks{Present Address: University of Yamanashi, Kofu, Yamanashi 400-8510, Japan}
\affiliation{\MCES}
\author{H. Hosono}
\affiliation{\NIMS}\affiliation{\MDXES}
\author{R.~Kadono}\thanks{email: ryosuke.kadono@kek.jp}\affiliation{\IMSS}

\date{\today}

\begin{abstract}
We performed muon spin rotation and relaxation ($\mu$SR) experiments to investigate the microscopic mechanism behind the high ionic conductivity ($\sigma$) exhibited by hydride (H$^-$) ions in lanthanum hydroxide LaH$_{3-2x}$O$_x$. The $\mu$SR spectra observed at 5--300 K in a sample with $x\approx0.25$ consist primarily of two components which are attributed to muons occupying tetrahedral (Tet) and octahedral (Oct) sites common to H$^-$. The spectra also indicate that muons at the Oct sites (Mu$_{\rm O}$) appear nearly stationary in the time scale of $\mu$SR ($\sim$10$^{-5}$ s), whereas those at the Tet sites (Mu$_{\rm T}$) are subject to the fluctuating local fields. The cusp-like peak in the fluctuation rate around 160 K and the decrease in linewidth at higher temperatures probed by Mu$_{\rm T}$ suggest that the jump motion of both Mu$_{\rm T}$ (via the vacant Oct sites) and surrounding Oct-site H$^-$ contributes to spin relaxation and that the fluctuation frequency is widely distributed. These results indicate that the implanted Mu behave as Mu$^-$ and that the jump motion of Mu$^-$/H$^-$ is restricted by the availability of nearby vacant sites. On the other hand, the activation energy for the jump is estimated to be 0.11(3) eV, which is significantly different from $\sim$1.3 eV evaluated from the temperature dependence of $\sigma$ at high temperatures ($\gtrsim400$ K). In our attempt  to resolve this discrepancy, we discuss problems inherent in interpreting $\sigma$ using the Arrhenius equation, and demonstrate that the behavior of H$^-$ ions can be better explained as a viscous fluid exhibiting a glass transition.
\end{abstract}

\maketitle


\section{Introduction}
\begin{figure}[t]
\includegraphics[width=0.9\linewidth,clip]{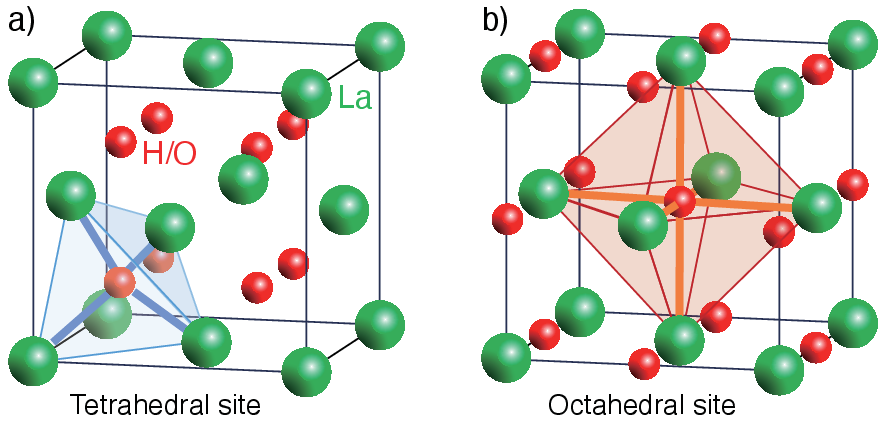}
	\caption{Schematic crystal structures of LaH$_{3-y}$ for (a) $y=1$ (fluorite structure) where H occupies tetrahedral (Tet) sites, and (b) the octahedral sites to be occupied by H for $y<1$ in addition to the Tet sites (not shown). Thus, the $\alpha$-BiF$_3$ structure with $y < 1$ can be regarded as a structure where the fluorite structure (a) and the rock salt structure (b) mutually interpenetrate.  In the case of LaH$_{3-2x}$O$_x$, oxygen preferentially occupies the Tet site.}
	\label{strc}
\end{figure}

Lanthanum hydride (LaH$_{3-y}$) has been intensively studied as a metal hydride exhibiting a metal-insulator transition~\cite{Fukai:05} or as an ``electride'' possessing catalytic functions~\cite{Hosono:21}.  The crystal structure of LaH$_{3-y}$ is based on the fluorite structure, with the cation lanthanum (La$^{3+}$) forming a face-centered cubic (fcc) sublattice (see Fig.~\ref{strc}). Since the hydride ion (H$^{-}$) occupies tetrahedral (Tet) interstitial sites of the La lattice at $y\approx1$, it is inferred that the excess electrons occupy octahedral (Oct) interstitial sites, forming a typical electron-rich structure that is highly receptive to H$^{-}$ ions. Indeed, for $y < 1$, hydrogen occupies the Oct sites as H$^{-}$ ions, ultimately approaching the $\alpha$-BiF$_3$ structure ($y \rightarrow 0$). This stoichiometric instability of the H$^{-}$ ions creates defects (vacancies) within the lattice, providing ``pathways'' for the fast diffusive motion of ions \cite{Majer:99}.

In recent years, LaH$_{3-y}$ has garnered renewed attention as one of the parent compounds in search for fast H$^{-}$ ion conductors~\cite{Kobayashi:16,Fukui:19,Fukui:22,Izumi:23, Sun:23}.  It was demonstrated that lanthanum hydroxides (LaH$_{3-2x}$O$_x$), where part of H$^{-}$ ions is substituted with O$^{2-}$ ions, exhibit a high conductivity of $2.6\times10^{-2}$ S/cm in the medium-temperature region (615 K) at $x=0.24$~\cite{Fukui:19}. More recently, it was reported that at room temperature ($\sim$298 K) for $x < 0.24$, an astonishing ionic conductivity of approximately $10^{-3}$ S/cm was achieved~\cite{Fukui:22}, which is more than $10^3$ times higher than previous values. This conductivity rivals that of high-speed proton (H$^+$) solid electrolytes, a leading candidate for next-generation batteries, dramatically increasing attention in the field of hydride ionics.

Similar to parent compounds, the high conductivity of H$^-$ ions in hydroxides is closely related to intrinsic defects present in the crystal structure. Moreover, O$^{2-}$ replacing H$^-$ at the Tet site creates H$^-$ ion vacancies at the Oct site.  For example, when $x \approx 0.25$, on average one out of every four Oct sites is secured to be vacant, allowing H$^-$ ions to temporarily reside there.  Note that such vacancies are necessary for H$^-$ ions to exhibit diffusive motion, as they have a relatively large ionic radius (1.2--1.5 \AA) comparable with that of O$^{2-}$. The high polarizability of H$^-$ ions \cite{Pauling:27} is also presumed to play a crucial role in lowering the energy barrier during diffusion, creating an environment in favor of rapid H$^-$ diffusion.

However, it turns out that the activation energy obtained by analyzing the temperature dependence of ionic conductivity using the Arrhenius equation is extremely high; it is $\sim$1.3 eV for $x=0.24$ \cite{Fukui:19}. Even recent results for $x < 0.24$ show values of 0.3--0.4 eV  \cite{Fukui:22}. While these values seem to be explained by a large-scale molecular dynamics (MD) simulations using neural network potentials (NNP) based on density functional theory (DFT) \cite{Fukui:22,Iskandarov:22}, they do not match the predicted activation energy of $\sim$0.15 eV for the elemental Tet-to-Oct site jump process by first-principles DFT calculations \cite{Fukui:19}.  Including this point, much of the microscopic mechanism underlying the high ionic conductivity exhibited by LaH$_{3-2x}$O$_x$ remains speculative.

To gain further insights into the H$^-$ ion conduction from the microscopic viewpoint, we performed muon spin rotation/relaxation ($\mu$SR) experiments in LaH$_{3-2x}$O$_x$ with $x\approx0.25$. $\mu$SR is one of the techniques used to ``simulate'' the behavior of H by implanting a positively charged muon ($\mu^+$) into the target material. While $\mu^+$ is an unstable subatomic particle studied in particle physics, it behaves as a light radioactive isotope of proton (possessing one-ninth the mass of a proton) when incorporated into matter. Therefore, to clarify this point, we will denote it as Mu$^+$ using an element symbol ``Mu"~\cite{Kadono:24a}.

As is shown below, the $\mu$SR results obtained at low temperatures (5--300 K) on a sample with $x\approx0.25$ suggest that the implanted $\mu^+$ indeed behaves as pseudo-H within the host, occupying both the Tet and Oct sites in the H$^-$-like state (Mu$^-$). Furthermore, while Mu$^-$ at the Oct site (Mu$^-_{\rm O}$) is nearly stationary, Mu$^-$ at the Tet site (Mu$^-_{\rm T}$) senses fluctuations in the internal magnetic field caused by jump motions of both Mu$^-_{\rm T}$ and surrounding H$^-$ ions at Oct sites. A detailed analysis of the $\mu$SR spectra subsequently performed using a spin relaxation model that explicitly accounts for surrounding ion jump motion with broad distribution of jumping frequency yielded an activation energy of 0.11(3) eV for the Tet-to-Oct site jump motion of Mu$^-_{\rm T}$. Based on these results, we address issues regarding conventional interpretation of ion conductivity based on thermally activated jump diffusion, and demonstrate that the ion conductivity is better explained by an alternative scenario of H$^-$ ions as a viscous fluid exhibiting a glass transition.

\section{$\mu$SR Experiment}

\par
The polycrystalline powder samples of LaH$_{2.5}$O$_{0.25}$ were obtained in a form of fractured disk after heating treatment under high-pressure, where the details of the synthesis are described elsewhere~\cite{Fukui:19}. Thermal properties were investigated  for part of the samples using differential scanning calorimetry (DSC) with a TA Instruments DSC Q2000. The sample pieces were mounted to the cold finger of the He gas-flow cryostat covering a cross section of $\sim$8 mm diameter (with $\sim$1 mm thick) against $\mu^+$ beam, and loaded to the S1 instrument ($\mu$SR spectrometer ARTEMIS) in the Materials and Life Science Experiment Facility (MLF), J-PARC, Japan. To prevent degradation of aerophobic specimen due to exposure to air, sample handling was performed inside a glove bag filled with flowing helium gas. Conventional $\mu$SR measurements were performed using a high-flux pulsed muon beam ($\approx10^3$ $\mu^+$s per pulse, with a repetition rate 25 Hz) transported to the S1 instrument. The ``flypast'' setup was adopted to minimize the background from $\mu^+$s that missed the sample to stop in the surrounding parts of the cryostat and/or spectrometer. The background yield versus sample diameter was calibrated using holmium slabs in the additional measurements conducted under the similar experimental conditions. The $\mu$SR spectra [the time-dependent decay-positron asymmetry, $A(t)$] which reflects the distribution of internal magnetic field at the $\mu^+$ site, was measured under a zero or longitudinal field (ZF/LF, parallel to the initial Mu polarization ${\bm P}_\mu$) at 50--320 K, and a weak transverse field (TF, perpendicular to ${\bm P}_\mu$), and were analyzed by the  `musrfit' codes based on the least-squares curve fit~\cite{musrfit}. 

\section{DATA ANALYSIS}\label{DatAna}
It was found in the preliminary analysis that the Lorentzian Kubo-Toyabe (LKT) function is needed in addition to the conventional Gaussian Kubo-Toyabe (GKT) function to reproduce the $\mu$SR spectra observed in the LaH$_{2.5}$O$_{0.25}$ sample. This suggests that the internal magnetic field experienced by Mu (the vector sum of dipole fields from randomly oriented magnetic moments of nearby $^{139}$La and $^1$H nuclei) exhibits strong disorder characteristic of amorphous glass. However, it is known that the dynamical LKT function, which usually incorporates the effects of internal magnetic field fluctuations using a strong collision model (as in the case for the musrfit~\cite{musrfit}), has a fundamental problem that it cannot describe the motional narrowing effect as it stands. Moreover, in such disordered systems, the fluctuation frequency generally exhibits a broad distribution, presenting the problem that it cannot be described by a single time constant~\cite{Uemura:85}. We resolved these issues based on the recently developed spin relaxation model that allows distinguishing the jump motion of muons themselves from that of surrounding ions \cite{Ito:24}, whose details are found in Appendices A and B.

\begin{figure}
\includegraphics[width=0.95\linewidth,clip]{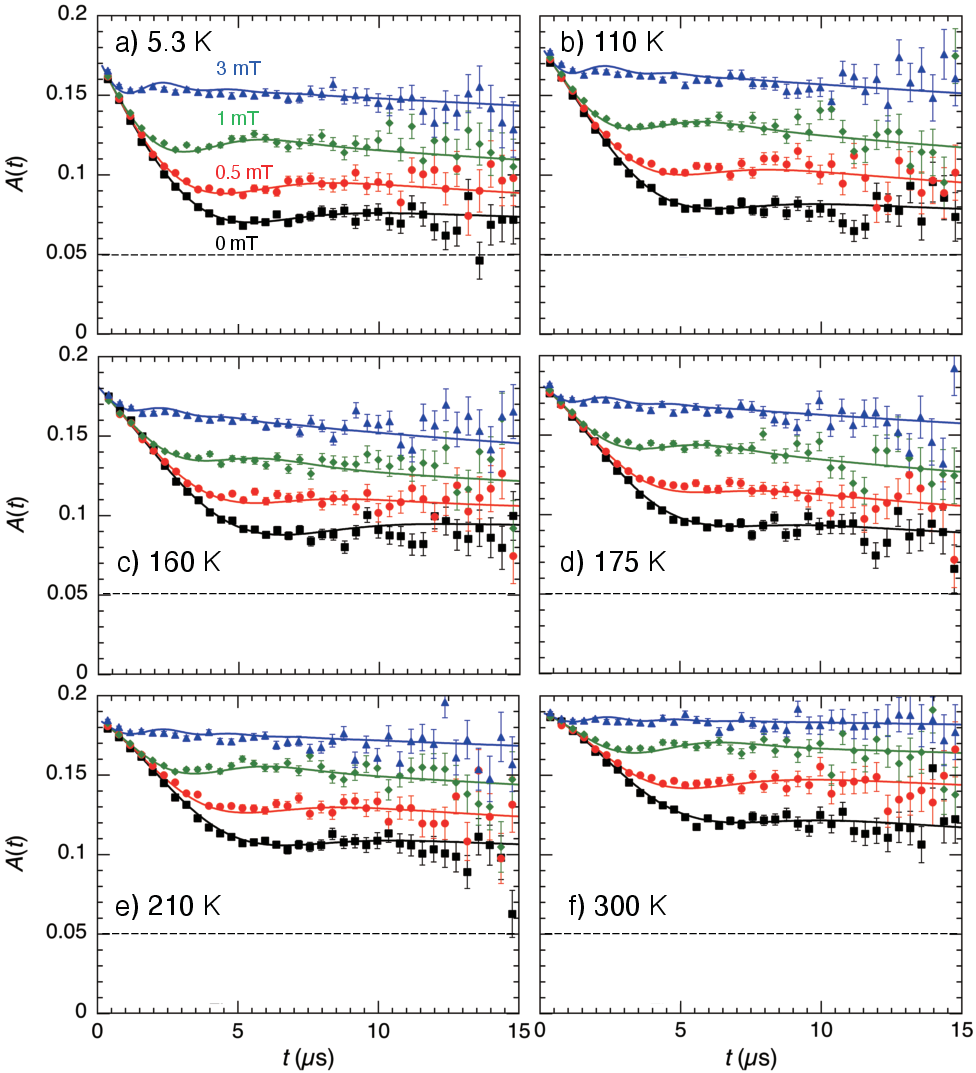}
	\caption{$\mu$SR time spectra observed in LaH$_{2.5}$O$_{0.25}$ at (a) 5.3 K, (b) 110 K, (c) 160 K, (d) 175 K,  (e) 210 K, and (f) 300 K, where the solid curves are results of global fits using Eq.~(\ref{asys}) with common parameters through four spectra ($\mu_0H_{\rm LF} = 0$, 0.5, 1, and 3 mT) at each temperature. Dashed lines indicate the background level ($A_{\rm b}\approx0.05$) estimated from calibration measurements. }
	\label{tspec}
\end{figure}

In the actual analysis, we consistently reproduced all observed spectra by assuming a recursive function of the following form:
\begin{equation}
A(t)\approx A_0[\rho g_z^{\rm G}(t;\Delta,H_{\rm LF})+(1-\rho)G_z^{\rm EA}(t;H_{\rm LF})]+A_{\rm b},\label{asyt}
\end{equation}
where $A_0$ is the partial asymmetry for the signal from the sample, $g_z^{\rm G}(t)$ is the static GKT function with $\Delta$ being the linewidth, $G_z^{\rm EA}(t;H_{\rm LF})\equiv G_z^{\rm EA}(t;\Lambda,\nu_{\rm f},\Phi,H_{\rm LF})$ is the extended LKT function incorporating the Edwards-Anderson (EA) parameter $\Phi$ (each defined by Eqs.~(\ref{gkt}) and (\ref{gea}) in Appendices which are extended to incorporate the effect of longitudinal magnetic field $H_{\rm LF}$), $\rho$ is the fractional yield for the static GKT function, and $A_{\rm b}$ is the background from $\mu^+$s stopped in the surrounding materials.  Eq.~(\ref{asyt}) was further approximated as follows for the practical implementation to curve fits,
\begin{equation}
A(t)\approx A_1[fg_z^{\rm G}(t;\Delta,H_{\rm LF})+(1-f)G_z^{\rm L}(t;\Lambda_{\rm eff},\nu_{\rm eff},H_{\rm LF})]+A_2,\label{asys}
\end{equation}
where $A_1$  denotes the partial asymmetry exhibiting depolarization, $A_2$ is the non-relaxation component, $G_z^{\rm L}(t)$ is the dynamical LKT function defined by Eq.~(\ref{Gdyn}) with $\Lambda_{\rm eff}$ and $\nu_{\rm eff}$ respectively being the effective linewidth and its fluctuation rate, and $f$ is the fractional yield of the GKT function. As is discussed below, this approximation results in a complex relationship between the physical parameters in $G_z^{\rm EA}(t)$ and those in $G_z^{\rm L}(t)$ that also depends on $\Phi$. (Since the two non-relaxation components $A_0(1-\rho)\zeta$ and $A_{\rm b}$ in Eq.~(\ref{asyt}) cannot be distinguished in the curve fitting, they are collectively defined as $A_2$.) Typical examples of the $\mu$SR asymmetry spectra analyzed by Eq.~(\ref{asys}) are shown in Fig.~\ref{tspec}, where the instrumental asymmetry was corrected using weak TF-$\mu$SR data. The spectra are reasonably reproduced by these fits including their LF dependence.

At this stage, several features of the temperature evolution of the $\mu$SR spectra can be noted. First, the quasi-static relaxation rate, determined by depolarization at the early part of the spectra (0--5 $\mu$s), decreases with increasing temperature. On the other hand, the full recovery of polarization due to weak LF observed at all temperatures indicates that the fluctuation frequency $\nu_{\rm eff}$ assumed in the LKT function component is lower than $\Lambda_{\rm eff}$. Furthermore, as most clearly seen in the ZF-$\mu$SR spectrum at 300 K, a non-relaxing component intrinsic to the sample exists, and its partial asymmetry increases with increasing temperature.  These features are discussed in more detail in Sect.~\ref{Rslt} based on the temperature dependence of the parameters obtained from curve fits.

\section{RESULT}\label{Rslt}

\begin{figure*}
\includegraphics[width=0.85\linewidth,clip]{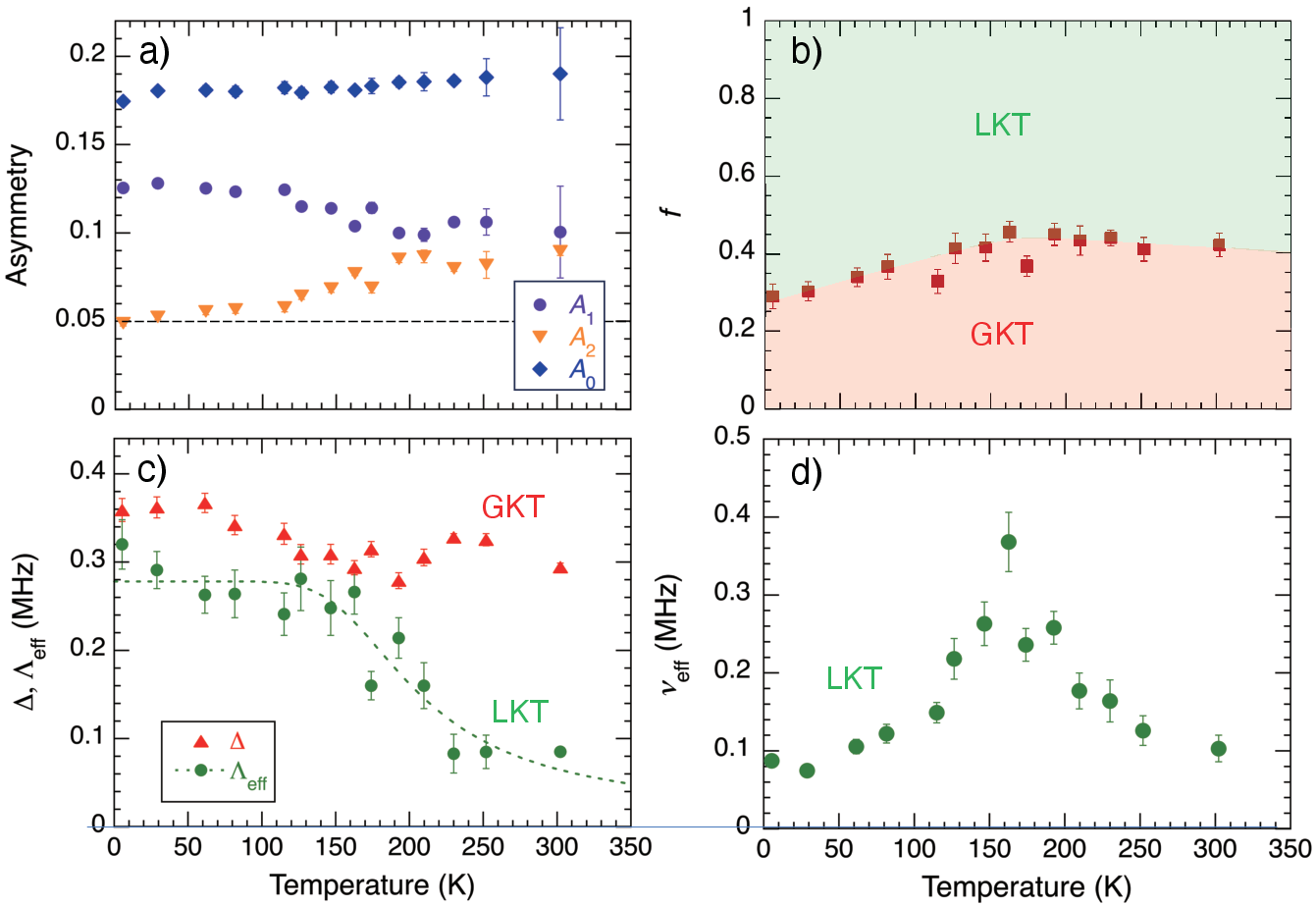}
	\caption{Temperature dependence of the parameters in Eq.~(\ref{asys}) deduced by curve fits: (a) Partial asymmetry $A_1$ for the component reprodiced by the sum of static GKT and dynamical LKT functions, $A_2$ for that with no relaxation, and the total asymmetry $A_0$ ($=A_1+A_2$). (b) The fractional yield for the component reproduced by the static GKT function (the rest represented by the dynamical LKT function). (c) The static linewidth for the GKT function ($\Delta$) and for the dynamical LKT function ($\Lambda_{\rm eff}$). (d) The fluctuation rate $\nu_{\rm eff}$ of the internal magnetic field for the dynamical LKT function. The dashed curve in (c) is the best fit with Eq.~(\ref{QT}). }
	\label{params}
\end{figure*}

The temperature dependence of the parameters in Eq.~(\ref{asys}) derived from curve fitting is summarized in Fig.~\ref{params}. A common feature observed in these six panels is that they show specific changes around the temperature $T_{\rm d}\sim160$ K  where $\nu_{\rm eff}$ in Fig.~\ref{params}(d) exhibits a cusp-like peak. In Fig.~\ref{params}(a), the partial asymmetry $A_1$ representing the relaxing component decreases from around $T_{\rm d}$, while $A_2$ of the non-relaxation component increases. In contrast, the total asymmetry ($A_0=A_1+A_2$) remains nearly constant, decreasing only slightly from 0.188(1) to 0.174(3) with decreasing temperature, suggesting a trade-off relationship between $A_1$ and $A_2$. The missing asymmetry is attributed to the formation of paramagnetic Mu commonly found in insulators~\cite{Kadono:24a}, from which the diamagnetic Mu (Mu$^+$ or Mu$^-$) state is readily distinguished by applying LF of a few mT (although the distinction between the two charge states are difficult).  The fractional yield $f$ of the GKT function component in Fig.~\ref{params}(b) gradually increases with temperatures up to $T_{\rm d}$ and remains nearly constant above $T_{\rm d}$. Regarding the linewidth in Fig.~\ref{params}(c), $\Delta$ decreases slightly from low temperatures to $T_{\rm d}$ and remains nearly constant above $T_{\rm d}$, whereas $\Lambda_{\rm eff}$  decreases significantly above $T_{\rm d}$.

These features can be understood within the framework of the phenomenological model underlying Eq.~(\ref{asys}). As the first step, let us discuss the correspondence between the two components and the Mu sites.  Mu as pseudo-H is expected to occupy the sites common to H. In the case of LaH$_{2.5}$O$_{0.25}$, the eight Tet sites within the unit cell (La$_4$H$_{10}$O$_{1}$) are fully occupied by H/O (see Fig.~\ref{strc}(a)), but Mu can occupy one of the four Oct sites that are, on average, vacant. Furthermore, considering that H has an exceptionally light mass, making it highly probable to be pushed out of Tet sites via direct knock-on processes during Mu implantation, it is expected that both sites are occupied randomly at the lowest temperature \cite{Kadono:24a}. The expected linewidth $\Delta$ at each site are calculated using Eq.~(\ref{delta_n}) in Appendix A and the known crystal structure \cite{Tamatsukuri:23} for two patterns where H and O (and La) have different coordination and results are shown in Table \ref{dlt}. (Note that our specific approach for the LKT function allows to assume $\Lambda\approx\Delta$: see Appendix A for more detail.)
 
\begin{table}[b]
\begin{tabular}{ccc}
\hline\hline
Configuration & $\Delta_{\rm T}$ (MHz) & $\Delta_{\rm O}$ (MHz) \\
 \hline
Pattern A (H:8/O:0@Tet, H:2/O:1@Oct) & 0.3019 & 0.3613 \\
Pattern B (H:7/O:1@Tet, H:3/O:0@Oct) & 0.3245 & 0.3456 \\
\hline\hline
\end{tabular}
\caption{Gaussian linewidth for either Mu$^-_{\rm T}$ ($\Delta_{\rm T}$) or Mu$^-_{\rm O}$ ($\Delta_{\rm O}$) calculated for the two H/O configurations. Pattern A: 8 Tet sites and 2 Oct sites occupied by H (1 Oct site occupied by O). Pattern B: 7 Tet sites and 3 Oct sites occupied by H (1 Tet site occupied by O). $\Delta_{\rm T}$ is for the Mu substituting one of H at the Tet site, while $\Delta_{\rm O}$ is for the Mu occupying the vacant Oct site.}\label{dlt}
\end{table}

One reason the values of $\Delta$ do not differ significantly in these four cases is that the nuclear magnetic moment of $^{139}$La  ($=2.778\mu_{\rm N}$, with $\mu_{\rm N}$ being the nuclear magneton) is nearly equal to that of $^1$H ($=2.793\mu_{\rm N}$). On the other hand, in both patterns A and B, the Oct-site values are larger than those of the Tet site. Assuming that these values correspond to $\Delta$ and $\Lambda_{\rm eff}$ in the static limit, comparing this systematic trend with the experimental result in Fig.~\ref{params}(c) suggests that the signal corresponding to the static GKT function originates from the Mu$^-_{\rm O}$, while that corresponding to the dynamical LKT function originates from the Mu$^-_{\rm T}$. This is further supported by the fact that the relative ratio of GKT-LKT components at the lowest temperature ($f:1-f\approx1:2$) is close to the ratio of Oct to Tet site numbers. The reason that $n({\bm H})$ at the Tet site becomes Lorentzian may be attributed to the strong disorder of H position at the Oct sites: a similar behavior is found in the $\mu$SR spectra observed in H-charged amorphous InGaZnO films~\cite{Kojima:19}.

The slight decrease of $\Delta$ above $T_{\rm d}$ may suggest the decrease of the average H occupancy at the Oct sites at higher temperatures. Meanwhile, the decrease of $\Lambda_{\rm eff}$ cannot be explained by such change in the linewidth itself, because it is reduced far below the static linewidth towards 300 K (see Sect.~\ref{DcnB}). It is also noticeable that the change accompanies the peculiar temperature dependence of $\nu_{\rm eff}$ and the increase of $A_2$.  Among these, the behavior of $\Lambda_{\rm eff}$ and $A_2$ can be immediately understood as corresponding to an increase in the EA parameter $\Phi$ (the contribution of the fast fluctuation component, leading to a decrease in $\Lambda_{\rm eff}=\sqrt{1-\Phi}\Lambda$) and the resulting increase in the relative ratio $\zeta$ representing the motional-narrowing component in Eq.~(\ref{gea}). Regarding the latter, Table \ref{dlt} demonstrates that quasi-static Mu$^-$ exhibits finite spin relaxation at any of these site-ion configurations, supporting the assumption that $\zeta$ corresponds to the relative yield of Mu$^-_{\rm T}$ in the motional-narrowing limit. (Note that similar examples in which Mu$^-$ states exhibit fast diffusion have been found in other materials \cite{Kadono:24a}.) 

The cusp-like peak in temperature dependence of $\nu_{\rm eff}$ in Fig.~\ref{params}(d) comprises evidence for the {\sl partial} fluctuation of the local field ${\bm H}(t)$ above $T_{\rm d}$ (corresponding to the situation that  $\Phi<1$ in Eq.~(\ref{AcfEA})). As discussed in detail in Appendix B, this peak is an apparent behavior resulting from approximating $G_z^{\rm EA}(t;\nu_{\rm f})$ in Eq.~(\ref{asyt}) by the LKT function $G_z^{\rm L}(t;\Lambda_{\rm eff},\nu_{\rm eff})$ in Eq.~(\ref{asys}) (see also Eq.~(\ref{gea})). Specifically, $G_z^{\rm EA}(t)$ is nearly equivalent with $G_z^{\rm L}(t;\Lambda,\nu_{\rm f})$ in the motional broadening regime ($\nu_{\rm f}\lesssim\Lambda$), with $\nu_{\rm eff}$ ($\propto\nu_{\rm f}$) showing a gradual increase with temperature.  As the crossover occurs to the motional narrowing regime ($\nu_{\rm f}\gg\Lambda$) with increasing $\nu_{\rm f}$, $G_z^{\rm EA}(t)$ evolves into another quasistatic limit (exhibiting slow relaxation $\propto2\Lambda_{\rm eff}^2/\nu_{\rm f}$ and the recovery of the 1/3 tail) with $\Lambda_{\rm eff}$ approaching $\sqrt{1-\Phi}\Lambda$. Consequently, in the curve fit, $G_z^{\rm L}(t;\Lambda_{\rm eff},\nu_{\rm eff})$ attempts to capture the change in the lineshape by a decrease in $\nu_{\rm eff}$ (with varying amplitude $1-\zeta$), while the fast fluctuating $\Phi$ component is incorporated into the non-relaxing term ($\propto\zeta$) in Eq.~(\ref{asys}).  Note that no such peak occurs for $\Phi=1$, where $\nu_{\rm eff}=\nu_{\rm f}$.

Now, considering that the self-diffusion of Mu$^-_{\rm T}$ should cause the full fluctuation of ${\bm H}(t)$ (corresponding to $\Phi=1$), the peak behavior of $\nu_{\rm eff}$ is uniquely attributed to the local motion of H$^-_{\rm O}$ around Mu$^-_{\rm T}$. Conversely, for the  motion of H$^-_{\rm O}$ to induce the fluctuation of ${\bm H}(t)$ corresponding to $\Phi= 1$, all the nearest neighboring (nn) H$^-_{\rm O}$ must jump coherently at once to comprise the single fluctuation event of  ${\bm H}(t)$, which is unlikely to occur (except at extremely high temperatures).  This conclusion is also supported by the similar behavior of fluctuation rate in hybrid perovskites deduced by analyzing $\mu$SR data using the conventional GKT function, where random rotational motion of cation molecules causes the partial fluctuation of ${\bm H}$ at the Mu site in the perovskite PbI$_3$ lattice~\cite{Koda:22,Hiraishi:23,Papadopoulos:24,Ito:24}. 

To quantitatively evaluate the temperature dependence of $\Lambda_{\rm eff}$, curve fitting was performed using the following empirical model for thermally activated two-state transition between the quasistatic and dynamical states,
\begin{eqnarray}
\Lambda_{\rm eff}&=&\sqrt{1-\Phi}\Lambda,\nonumber\\ 
\Phi  &=& \frac{1}{1+N^{-1}\exp\left(\frac{E_{\rm a}}{k_BT}\right)},\label{QT}
\end{eqnarray}
where $N$ refers to the parameter related with the density of states, and $E_{\rm a}$ is the activation energy  \cite{Cox:06b}, the results of which are shown as a dashed line in Fig.~\ref{params}(c). From this, it can be seen that Eq.~(\ref{QT}) reasonably reproduces the temperature dependence of $\Lambda_{\rm eff}$ with $\Lambda=0.278(13)$ MHz, $N= 1.3(2.1)\times10^3$, and $E_{\rm a}=0.11(3)$ eV. The activation energy is close to 0.101(9)  eV estimated for that of the Tet-Oct transition in LaH$_{3-y}$ by quasielastic neutron scattering  (QENS) as well as the migration barrier ($\sim$0.11 eV) obtained from DFT-NNP-based MD simulations~\cite{Tamatsukuri:23}. (These values are also consistent with $E_{\rm a}=0.150(10)$ eV for long-range H$^-$ diffusion in LaH$_{3-y}$ inferred from the previous QENS experiment~\cite{Udovic:99}.)  
Thus, while our results do not directly provide the fluctuation rate $\nu_{\rm f}$, the magnitude of $E_{\rm a}$ strongly suggests that the local Mu/H dynamics is dominated by the Tet-Oct jump motion in the relevant temperature range. 

\section{DISCUSSION}\label{Dcn}
\subsection{Charged state of diamagnetic Mu}
In the previous Sect.~\ref{Rslt}, we attributed the charged state of diamagnetic Mu to negatively charged states (Mu$_{\rm T}^-$ and Mu$_{\rm O}^-$). This is based on the fact that it occupies the same site as H$^-$. Furthermore, although only with a small yield, we showed that some Mu adopts a paramagnetic state (Mu$^0$, observed as a missing asymmetry) at low temperatures. The reason Mu takes these two distinct charged states is explained by the ``ambipolarity model,'' recently proposed to achieve a unified understanding of Mu's local electronic state in insulators/semiconductors \cite{Kadono:24a}. This model suggests the simultaneous formation of donor-like and acceptor-like relaxed-excited states due to local electronic excitation accompanying the kinetic energy of injected $\mu^+$. The presence of two distinct charged Mu states in electronically insulating LaH$_{3-2x}$O$_x$ can be also understood within the framework of this model.

Until very recently, diamagnetic Mu has been often interpreted simply as Mu$^+$, regardless of the electronic properties of host materials. This simplification was particularly unavoidable in hydride materials, where the coexistence of Mu with abundant H atoms further complicated the interpretation of experimental results. It is worth mentioned that recent advances in DFT calculations (including those focused on Mu \cite{Blundell:23}) and the ambipolarity model based on them have revealed that the diamagnetic Mu in MgH$_2$ \cite{Sugiyama:19} and NaAlH$_4$ \cite{Kadono:08}, previously regarded as Mu$^+$, actually involves Mu$^-$ states \cite{Kadono:24a}.

\subsection{Strong correlation of diffusive motion between Mu and H}\label{DcnB}
In recent years, $\mu$SR studies on ion dynamics have been attracting considerable attention \cite{Sugiyama:09,Sugiyama:11,Sugiyama:12,Mansson:13,Baker:11,Umegaki:17,Sugiyama:20,Ohishi:22a,Ohishi:22b,Umegaki:22,Ohishi:23,Forslund:25,Tatara:25}. In these studies, it has been argued that the fluctuation rate $\nu$ of internal fields deduced by the analysis using dynamical GKT function stem solely from cation diffusion motion, based on the assumption that Mu is immobile (e.g., due to OMu bond formation in insulating oxides). However, it has been demonstrated that in such materials, the coexistence of immobile and mobile ions that give rise to {\sl partially} fluctuating internal fields does not allow such a naive interpretation of $\nu$ (see Appendix B) \cite{Ito:24}. Moreover, the evidence is accumulating that fluctuations due to self-diffusion of Mu are often dominant in these materials \cite{Ito:25c}. Under these circumstances, LaH$_{3-x}$O$_{x}$ provides a unique opportunity to specifically discuss the self-diffusion of Mu$^-$, the diffusion motion of surrounding H$^-$ ions, and their relationship to the behavior of the $\mu$SR spectra. Furthermore, since Mu and H occupy the same site as isotopes, it provides crucial information for microscopically understanding H$^-$ diffusion processes at high densities.

The present experimental results immediately suggest that the jump motion of Mu$^-$ is strongly rate-limited by that of H$^-$. The formation enthalpy indicates that the most stable site for Mu/H is the Tet site, implying that Mu$^-$ stopped at the Oct site is expected to immediately transition to the vacant Tet site (if available). Therefore, the observed quasi-static Mu$^-_{\rm O}$'s are interpreted as those not adjacent to such vacant Tet sites: the possibility to find them must be rare at low temperatures where the Tet sites are mostly occupied by H$^-$. On the other hand, while the jump motion of Mu$^-_{\rm T}$ is also rate-limited by the vacancy status of the Oct site, one out of four Oct sites is always vacant for La$_4$H$_{10}$O, allowing jump motion to occur via thermal activation. For long-range diffusion to progress, however, another vacancy must exist in the Tet site adjacent to the vacant Oct site.  This requires that a significant number of H$^-_{\rm T}$'s must be excited to the Oct site. In the temperature range $T>T_{\rm d}$ where this situation is suggested to arise, the motion of the nn H$^-_{\rm O}$'s may dominate spin relaxation of Mu$^-_{\rm T}$. Here, it must be noted that the fluctuating fields from the jumping $^1$H nuclei coexist with the quasi-static contributions from the $^{139}$La matrix with nearly equal amplitude, suggesting that $1-\Phi$ should approach $\sim$0.5 ($\Lambda_{\rm eff}/\Lambda=\sqrt{1-\Phi}\approx0.71$) as long as Mu$^-_{\rm T}$ remains immobile at high temperatures. The fact that $1-\Phi$ is actually reduced far below this at higher temperatures indicates that Mu$^-_{\rm T}$ also exhibits jump motion for $T>T_{\rm d}$. These interpretations are consistent with our model of spin relaxation shown in Sect.~\ref{DatAna}.

Regarding the potential influence of electrons (those generated by thermal ionization of H$^-$ or by the $\mu^+$-induced electronic excitations), the electronic conductivity of the sample with $x\approx0.25$  measured by dc current is less than $\sim$1$\times10^{-4}$ S/cm at 610 K \cite{Fukui:19}. This is more than two orders of magnitude smaller than the ionic conductivity at the corresponding temperature ($\sim$0.02 S/cm), suggesting that their contribution to the spin relaxation process (e.g., via charge/spin exchange with Mu) is negligible. 

\subsection{H$^-$ diffusion and glass transition}
As mentioned earlier, it has been reported that analyzing the ionic conductivity exhibited by LaH$_{3-2x}$O$_x$ with $x=0.24$ using the Arrhenius equation reveals a large activation energy $E_{\rm a}\sim1.3$ eV, an order of magnitude larger than that of the Tet-Oct jump. Furthermore, its pre-exponential factor ($A\sim$10$^{12}$ S/cm/K) is several orders of magnitude larger than previously reported values or those predicted by MD simulations. These anomalies have led to a scenario of introducing the associated enthalpy $H_{\rm assoc}$ arising from the lanthanum adjacent to oxygen effectively carrying a large positive charge. In addition, by assuming that  the associated enthalpy decreases with temperature as $H_{\rm assoc}=a-bT$, they suggested the large $A$ could be understood through the contribution of the second term ($A\rightarrow Ae^{b/k_B}$) ~\cite{Fukui:19}.

However, it must be noted that there is an upper limit to the temperature range where the diffusion process governing ion conduction is assumed to be described by the Arrhenius equation via over-barrier jumps,
\begin{equation}
D = \frac{1}{z}\nu_0 d^2\exp\left[-\frac{E_{\rm a}}{k_BT}\right],\label{Arrh}
\end{equation}
where $z$ is the number of sites accessible for single jump, $d$ is the jump distance, and $\nu_0$ is the attempt frequency for the jump~\cite{Fukai:05}. In diffusion via over-barrier jumps, $\nu_0$ is determined by the Debye temperature $\Theta$ in the solid (i.e., $\nu_0\approx k_B\Theta/h$  with $h$ being the Planck constant). Therefore, in the temperature region where the mean kinetic energy of ions is much greater than $k_B\Theta$, the residence time $\tau_{\rm r}$ for ions in the potential well can be neglected ($\tau_{\rm r}\ll\nu_0^{-1}$). Thus the diffusion process becomes that of fluid-like motion well described by the Einstein's relation~\cite{Einstein:05}, 
\begin{equation}
D = \mu k_BT, \label{Enstn}
\end{equation}
where $\mu$ is the ion mobility.  Given that the Debye temperature of LaH$_{3-y}$ determined by heat capacity measurements is 350--400 K~\cite{Kai:89}, such a fluid-like diffusion process may dominate in the temperature range where high ionic conductivity is observed.

\begin{figure}[t]
\includegraphics[width=0.9\linewidth,clip]{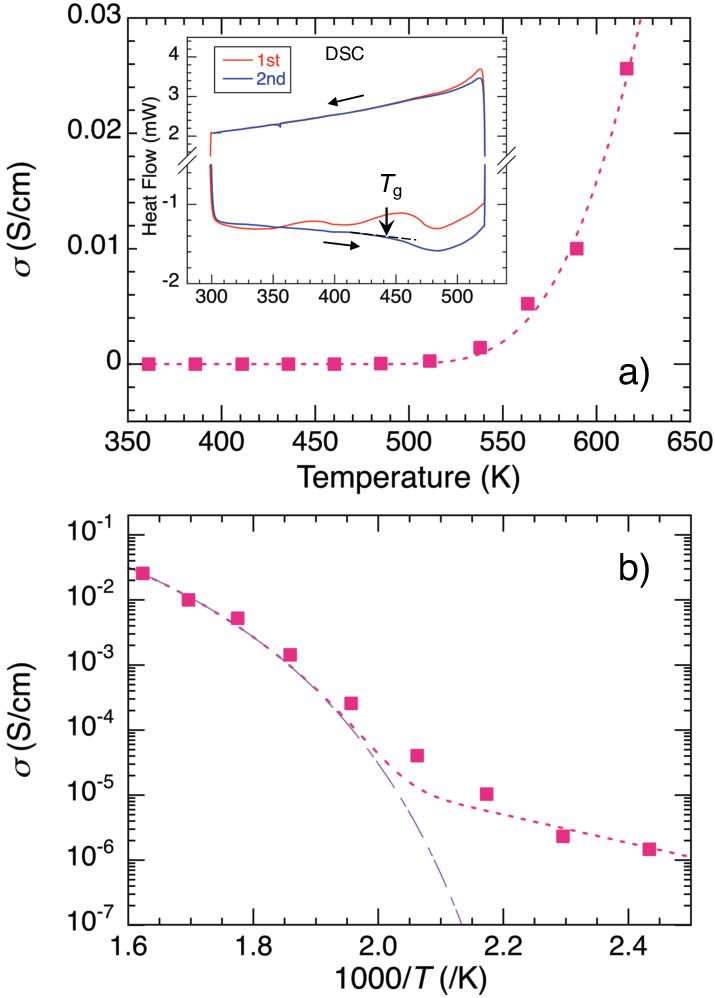}
	\caption{Ion conductivity of LaH$_{3-2x}$O$_x$ at $x=0.24$ plotted against $T$ on a linear scale (a) and on a logarithmic scale against $1/T$ (b) (after Ref.~\cite{Fukui:19}). Dashed lines are fits using Eqs.~(\ref{VFT}) and (\ref{JG}). Inset in (a): DSC curves measured for the present sample ($x\approx0.25$), where the onset of an endothermic reaction is inferred at $T_{\rm g}\approx440$ K in the second loop which is relatively free from the strain effect of the sample prepared by high pressure synthesis. The dashed line shows extrapolated baseline. }
	\label{sigma}
\end{figure}

Upon reexamining the reported temperature dependence of $\sigma$ at $x=0.24$ shown in Fig.~\ref{sigma}(a), it appears to increase sharply above a threshold temperature ($\sim$500 K). Furthermore, considering the disorder associated with the random occupation of Tet sites by oxygen that leads nearby H$^-_{\rm O}$ into off-center position~\cite{Fukui:19} (which is also in line with the strong tendency of OH formation in lanthanum oxides~\cite{Bernal:87}), it may be reasonable to interpret the behavior of $\sigma$ as a glass transition in the fluidic regime where mobility is described by Eq.~(\ref{Enstn}).  In fact, DSC measurements performed on the sample used for our $\mu$SR measurements ($x\approx0.25$) confirmed the onset of an endothermic reaction, as shown in the inset of Fig.~\ref{sigma}(a). This is likely attributable to the glass transition at $T_{\rm g}\approx440$ K. Note that there have been previous instances where a correlation between glass transition and fast cation conduction has been suggested~\cite{Kawamura:86,Oguni:94,Ngai:23}.

In the supercooled state of a glassy material, it is empirically known that the temperature dependence of viscosity can be approximately described by the Vogel-Fulcher-Tammann  equation,
\begin{equation}
\eta=\eta_0\exp\left[-\frac{B}{k_B(T-T_0)}\right], \:\:(T>T_0)\label{visc}
\end{equation}
where $B$ represents the energy parameter related with the fragility, and $T_0$ is the Vogel temperature (corresponding to the ideal glass transition temperature where the relaxation time $\tau$ and the barrier for viscous flow increase to infinity), which is slightly below the actual $T_{\rm g}$~\cite{Vogel:21,Fulcher:25,Tammann:26,Donth:01,Ngai:23}.  Eq.~(\ref{visc}) indicates that the effective activation energy for ion motion increases sharply with decreasing temperature as it approaches $T_0$. It is presumed that this very situation in the glass transition is the reason why a large activation energy is required when attempting to describe $\sigma$ using a simple Arrhenius equation.

Since viscosity is inversely proportional to $\tau$ ($=1/\nu$) of ion motion (i.e., $\nu\propto\eta$), we can assume the following relation,
\begin{equation}
\sigma\propto \frac{nq^2l^2}{k_BT\tau}=\frac{A}{T}\exp\left[-\frac{B}{k_B(T-T_0)}\right],\label{VFT}
\end{equation}
where $n$ is the ion density, $q$ is the ionic charge, and $l$ is the mean free path.  As shown in Fig.~\ref{sigma}(a), curve fit using Eq.~(\ref{VFT}) reasonably reproduces data at $x=0.24$ with $A=4.9\times10^3$ S/cm/K, $T_0=407$ K, and $B=0.104$ eV. (The similar fit applied to the data obtained for another sample with the nominally same composition (found in the Supplement of Ref.\cite{Fukui:19}) yielded $A=2.7\times10^4$ S/cm/K, $T_0=460$ K, and $B=0.089$ eV, in reasonable agreement with the above result.) It is noticeable that $T_0$ is close to $\Theta$ in LaH$_{3-y}$ for $0\le y\le1$~\cite{Kai:89}.

Interestingly, when $\sigma$ is plotted on a logarithmic scale against the inverse temperature (see Fig.~\ref{sigma}(b)), it can be seen that $\sigma$ begins to deviate from Eq.~(\ref{VFT}) below around $T_{\rm g}$. One possible cause for this is the distribution of $T_{\rm g}$ associated with the system's inhomogeneity. On the other hand, such branching of the relaxation rate with decreasing temperature (often referred to as ``decoupling'') is commonly observed in glass transitions and is called the $\beta$ process (or Johari-Goldstein process)~\cite{Kawamura:86,Ngai:23}. (In contrast, the fast relaxation process at higher temperatures is called the $\alpha$ process.) The $\beta$ process is well described by the Arrhenius equation, where its activation energy is much  larger than that of the elementary process of ion motion. This generally leads to the interpretation that the $\beta$ process corresponds to motion involving multiple ions or molecules. In fact, considering such an additional contribution
\begin{equation}
\sigma_\beta=\sigma_{\beta0}\exp\left[-\frac{E_\beta}{k_BT}\right]\label{JG}
\end{equation}
to Eq.~(\ref{VFT}) allows us to reproduce the low-temperature behavior as seen in Fig.~\ref{sigma}(b) with $\sigma_{\beta0}\approx$0.3 S/cm$^{-1}$ and an activation energy $E_\beta\approx$0.43 eV.

The activation energy $E_{\rm a}$ for the Tet-Oct jump motion of H$^-$ ions obtained in this work (and from the previous QENS experiments as well) is remarkably close to the value of parameter $B$ in Eq.~(\ref{VFT}), which is likely no coincidence. Generally, the fast relaxation processes occurring far above $T_{\rm g}$ are dominated by local ion motion driven by phonons. Since such motion is presumed to be close to that of the elementary process, this agreement is readily understood by considering that its energy scale governs $B$. Furthermore, the reason for the increase in the effective activation energy with decreasing temperature towards $T_0$  can be attributed to the development of correlation in the ion motion with surrounding ions.  DFT-NNP-based MD simulations of H$^-$ ion behavior revealed ``cooperative motion'' where multiple adjacent H$^-$ ions repel each other while migrating over long distances~\cite{Fukui:22,Iskandarov:22}. This may provide a microscopic picture of the $\beta$ process. Considering these points, it is concluded that the behavior of H$^-$ ion conductivity in LaH$_{2.5}$O$_{0.25}$ can be understood semi-quantitatively within the framework of viscous fluid exhibiting glass transition.

\section{Conclusion}
We have successfully derived a phenomenological spin relaxation function $G_z^{\rm EA}(t)$ that enables quantitative analysis of $\mu$SR spectra observed in LaH$_{2.5}$O$_{0.25}$, considering the Lorentzian distribution of the internal field ${\bm H}(t)$ and its partial fluctuations due to H$^-$ ion diffusion by introducing the Edwards-Anderson parameter $\Phi$ in the autocorrelation of ${\bm H}(t)$. We then showed that $G_z^{\rm EA}(t)$ was approximated by the Lorentzian Kubo-Toyabe function $G_z^{\rm L}(t;\Lambda,\nu_{\rm eff})$ (plus a non-relaxing contribution), where the monotonous increase in the fluctuation rate of ${\bm H}(t)$ is represented by the peak behavior of the effective fluctuation rate $\nu_{\rm eff}$ when the fluctuation is {\sl partial} ($\Phi<1$, whereas no such peak is expected for Mu diffusion corresponding to $\Phi=1$). As a result, we found that $\nu_{\rm eff}$ indeed exhibits the peak behavior, providing strong evidence for the jump motion of H$^-_{\rm O}$ around Mu$^-_{\rm T}$. In addition, the gradual onset of Mu$^-_{\rm T}$ diffusion was suggested from the reduction of $1-\Phi$ below $\sim$0.5 at high temperatures.  These results led us to conclude that the fundamental process of ion diffusion is the jump motion of Mu$^-$/H$^-$ ions at the Tet sites via Oct-site vacancies with an activation energy of 0.11(3) eV, consistent with conclusions obtained from preceding QENS experiments and DFT calculations for the Tet-to-Oct site jump process. Furthermore, we pointed out that the enormous activation energy obtained from conventional ion conductivity analysis may be an apparent value resulting from the assumption of thermally activated diffusion behavior in the temperature range where ion diffusion becomes fluid-like. Consequently, we demonstrated that the behavior of H$^-$ ions is better understood as that of a viscous fluid exhibiting a glass transition which is also consistent with the local motion of H$^-$ ions inferred from $\mu$SR experiment.

\begin{acknowledgments}
We would like to thank the MLF staff for their technical support during $\mu$SR experiment, which was conducted under the support of Inter-University-Research Programs by Institute of Materials Structure Science, KEK (Proposal No. 2013MS01). Thanks are also to T. Yamada for helping DSC measurements in the user laboratories at the Neutron Science and Technology Center, CROSS. This work was supported by the Elements Strategy Initiative to Form Core Research Centers, from the Ministry of Education, Culture, Sports, Science, and Technology of Japan (MEXT) under Grant No.~JPMXP0112101001, and partially by the MEXT Program: Data Creation and Utilization Type Material Research and Development Project under Grant No.~JPMXP1122683430. M.H. also acknowledges the support of JSPS KAKENHI Grant No.~22K05275 from MEXT.  
\end{acknowledgments}

\section*{Data Availability}
The DSC data in Fig.~\ref{sigma} inset are available upon reasonable request to the corresponding author.


\setcounter{table}{0}
\setcounter{equation}{0}
\renewcommand{\thetable}{A\arabic{table}}
\renewcommand{\theequation}{A\arabic{equation}}
\subsection*{APPENDICES}

\subsection{Dynamical model for the Lorentzian Kubo-Toyabe relaxation function}\label{dlkt}
In general, the spin relaxation of diamagnetic muons is mainly governed by the random local fields ${\bm H}$ exerted from the nuclear magnetic moments of the nearby atoms. In particular, since muon sites (${\bm r}_\mu$) in solid crystals occupy relatively symmetric interstitial positions so that several number of nuclear magnetic moments are located nearly equidistantly, the corresponding density distribution,
\begin{equation}
n({\bm H})=\int\delta({\bm H}- {\bm H}({\bm r}_\mu))d{\bm r}_\mu, \label{nb}
\end{equation}
is well approximated by the Gaussian distribution,
\begin{equation}
n(H_\alpha)=\frac{\gamma_{\mu}}{\sqrt{2\pi}\Delta}
\exp\left(-\frac{\gamma_{\mu}^{2}H_\alpha^{2}}{2\Delta^{2}}\right),  \:\:   (\alpha=x,y,z),
\label{ph}
\end{equation}
where with $\gamma_\mu=2\pi\times13.553$ kHz/Oe is the muon gyromagnetic ratio. Note that the mean value $E(H)$ and standard deviation $V(H)$ for Eq.~(\ref{ph}) are well defined: $E(H)=\int_{-\infty}^{\infty} H\:n(H)dH=0$, $V(H)=E([H-E(H)]^2)=\Delta^2/\gamma_\mu^2$. 
The magnitude of $\Delta$ (corresponding to the linewidth of the spin relaxation) is determined by nuclear magnetic moments exerting random local fields to Mu:
\begin{equation}
\Delta^2=\frac{1}{2}\gamma_\mu^2\gamma_I^2\sum_i\sum_{\alpha=x,y}\sum_{\beta=x,y,z}(\hat{A}_i^{\alpha\beta}{\bm I}_i)^2, \label{delta_n}
\end{equation}
with 
\begin{equation} 
(\hat{A}_i)^{\alpha\beta}
=\frac{1}{r^3_i}(\frac{3\alpha_i \beta_i}{r^2_i}-\delta_{\alpha\beta}) \:\:(\alpha,\beta=x,y,z)\label{diptensor}
\end{equation}
being the dipole tensor for the $i$th nuclear magnetic moment $\gamma_I{\bm I}_i$ situated at ${\bm r}_i$ from the Mu site. Since the magnetic dipolar field decays proportionally to $1/r_i^3$, the magnitude of $\Delta$ is dominated by the nn nuclear magnetic moments around Mu. The more precise value covering large $i$ is calculated using a self-made computer program. The corresponding relaxation function is obtained analytically to yield the  {\sl static} GKT function \cite{Kubo:66,Hayano:79},
\begin{equation}
g^{\rm G}_{z}(t;\Delta) =   \frac{1}{3}+\frac{2}{3}(1-\Delta^{2}t^{2})e^{-\frac{1}{2}\Delta^{2}t^{2}}, \label{gkt}
\end{equation}
where the ``1/3 tail'' corresponds to the probability that ${\bm H}({\bm r}_\mu)$ is parallel to ${\bm P}_\mu$.

\begin{figure}
\includegraphics[width=0.95\linewidth,clip]{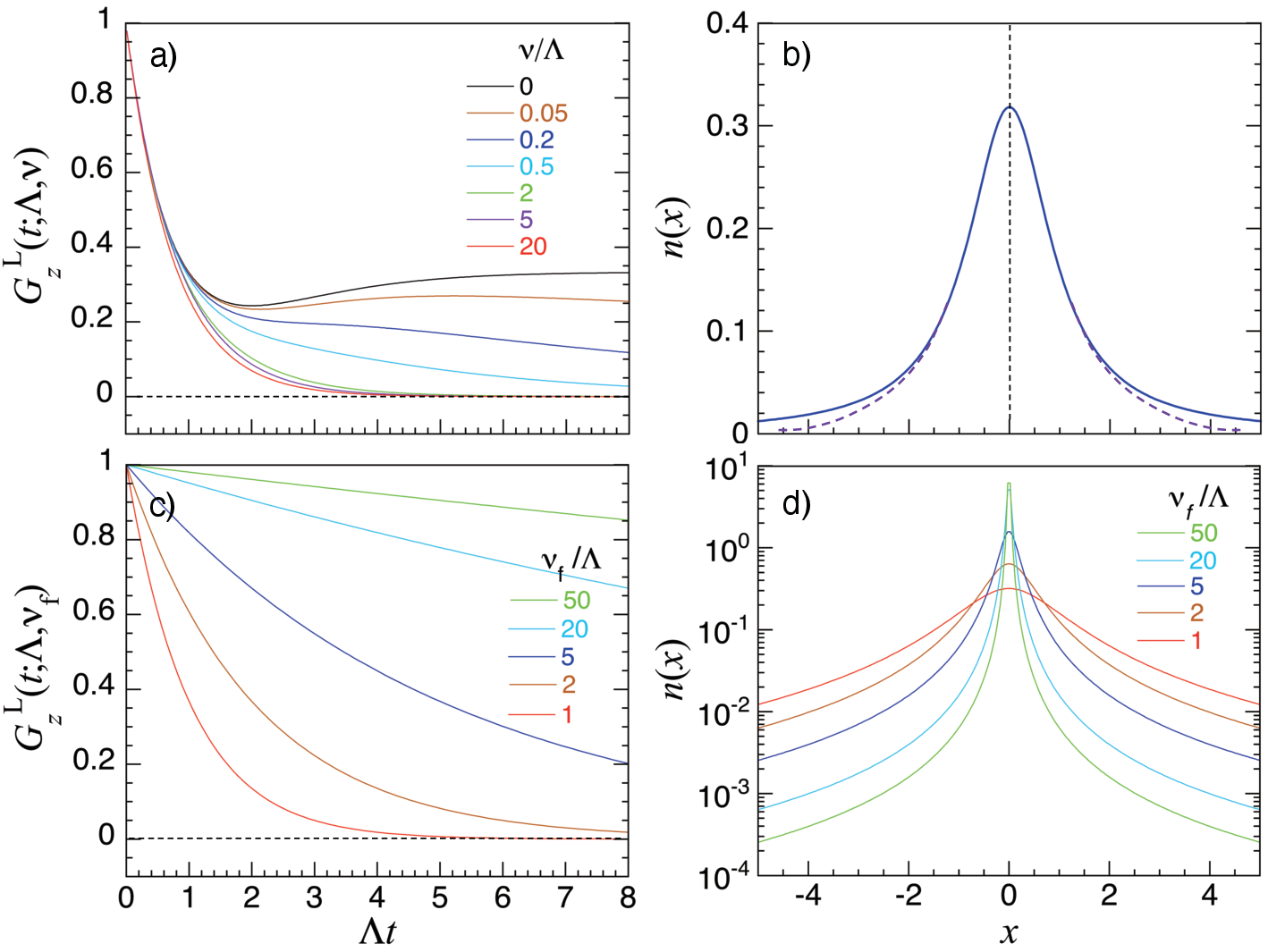}
	\caption{(a) Lorentzian Kubo-Toyabe (LKT) function under a zero external magnetic field, where the fluctuation of local field at a rate $\nu$ (normalized by the linewidth $\Lambda$)  is incorporated by strong collision approximation. (b) Lorentzian density distribution function, where $x=\gamma_\mu H/\Lambda$: dashed lines shows possible cutoffs to remove unphysical property (not considered in this study). (c) LKT function with $\Lambda$ replaced with $\Lambda'=2\Lambda^2/\nu_{\rm f}$, where $\nu_{\rm f}\gtrsim\Lambda$, and (d) the corresponding $n(x)$.}
	\label{LKTfun}
\end{figure}

On the other hand, for relatively disordered systems such as spin glass in magnetism, the density distribution is known to be better approximated by that of the Lorentzian  \cite{Uemura:85},
\begin{equation}
n(H_\alpha)=\frac{1}{\pi}\frac{\Lambda/\gamma_\mu}{(\Lambda/\gamma_\mu)^2+H_\alpha^2}, \:\:(\alpha=x,y,z)\label{nh}
\end{equation}
\begin{equation}
n(|{\bm H}|)=\frac{1}{\pi^2}\frac{\Lambda/\gamma_\mu}{((\Lambda/\gamma_\mu)^2+\gamma_\mu^2|{\bm H}|^2)^2}4\pi|{\bm H}|^2,\label{lkt}
\end{equation}
where $\Lambda/\gamma_\mu$ is a half-width at half maximum (HWHM).
The LKT functions for ZF and LF are derived from Eq.~(\ref{lkt}), and they are  conventionally used in the analysis as the corresponding phenomenological model. In particular, the relaxation function under ZF is given by a convenient analytical form \cite{Uemura:85}, 
\begin{equation}
g_z^{\rm L}(t;\Lambda)=\frac{1}{3}+\frac{2}{3}(1-\Lambda t)e^{-\Lambda t},\label{gzlt}
\end{equation}
which is also characterized by the $1/3$ term [see Fig.~\ref{LKTfun}(a)].  

However, as mentioned earlier, the LKT function has an unphysical property that it cannot reproduce the motional narrowing effect when time-dependent fluctuations of ${\bm H}$ must be taken into account \cite{Uemura:85}. Specifically, given that the fluctuation of ${\bm H}$ with a mean frequency $\nu$ is described by an autocorrelation function for the Gaussian-Markov stochastic process
\begin{equation}
C(t)=\frac{\langle {\bm H}(t){\bm H}(0)\rangle}{\langle H(0)^2\rangle}= \frac{\langle {\bm H}(t){\bm H}(0)\rangle}{\Delta^2/\gamma_\mu^2}=e^{-\nu t}, \label{Acf}
\end{equation}
with $\langle H(0)^2\rangle=\Delta^2/\gamma_\mu^2$ denoting the second moment of $n({\bm H})$,
and that the strong collision approximation can be applied to Eq.~(\ref{gzlt}),  the dynamical LKT function is obtained by solving the integral equation,
\begin{equation}
G^{\rm L}_z(t;\Lambda,\nu)=e^{-\nu t}g^{\rm L}_z(t;\Lambda)+\nu\int_0^t d\tau e^{-\nu(t-\tau)}g^{\rm L}_z(t-\tau;\Lambda)G^{\rm L}_z(\tau).\label{Gdyn}
\end{equation} 
As shown in Fig.~\ref{LKTfun}(a), the lineshape exhibits a decay of the 1/3 tail with increasing $\nu/\Lambda$, completely damped at around $\nu/\Lambda\approx1$ where $G^{\rm L}_z(t;\Lambda,\nu)\simeq\exp(-\Lambda t)$.  Meanwhile, it remains unchanged with further increase in $\nu$ to the limit of $\nu/\Lambda\rightarrow\infty$, which is unphysical in that it does not exhibit motional narrowing in the way predicted by the relation introduced by Bloemburgen, Purcell, and Pound (BPP) in the field of NMR \cite{Bloembergen:48}: 
\begin{equation}
\Lambda\approx\frac{2\Delta^2\nu}{\nu^2+\gamma_\mu^2H^2}.\label{bpp}
\end{equation}
This is because the mean value and standard deviation cannot be defined for the Lorentzian $n(H)$: $E(H)=\int_{-\infty}^{\infty} H\:n(H)dH=\infty$, $V(H)=E([H-E(H)]^2)=\infty$ [thus invalidating the implicit assumption that $\Lambda\approx\Delta$ in Eq.~(\ref{Acf})]. The origin of these anomalies is that $n(H)$ has finite weight even at $H\rightarrow\infty$, corresponding to the unphysical situation that the distance between Mu and nuclear magnetic moments [$r_i$ in Eq.~(\ref{diptensor})] can fall to zero ($H\propto r_i^{-3}\rightarrow\infty$ for $r_i\rightarrow0$). Therefore, it is categorically impossible for $H$ given by a Lorentz-type $n(H)$ to satisfy the narrowing condition that $\gamma_\mu H\ll\nu$.

This problem was first discussed in detail by Uemura and coworkers \cite{Uemura:85,Uemura:80}. They focused on the fact that in the static limit, the LKT function can be expressed as a superposition of GKT functions over the various linewidth $\Delta$ given by a Gaussian density distribution:
\begin{eqnarray}
g_z^{\rm L}(t;\Lambda)&=&\int_0^\infty g_z^{\rm G}(t;\Delta)\rho(\Delta)d\Delta,\label{GLconv}\\
\rho(\Delta)&=&\sqrt{\frac{2}{\pi}}\frac{\Lambda}{\Delta^2}\exp\left(-\frac{\Lambda^2}{2\Delta^2}\right).\nonumber
\end{eqnarray}
Based on this, they proposed the following approach: first, apply the strong collision model as a stochastic process to the GKT function with constant $\Delta$ to obtain the dynamical GKT function, 
\begin{equation}
G_z^{\rm G}(t;\Delta,\nu)=e^{-\nu t}\left[g_z^{\rm G}(t;\Delta)+\nu\int_0^t g_z^{\rm G}(t_1;\Delta)g_z^{\rm G}(t-t_1;\Delta)dt_1\right.\nonumber
\end{equation}
\vspace{-0.5cm}
\begin{equation}
\left.+\nu^2\int_0^t\int_0^{t_2} g_z^{\rm G}(t_1;\Delta)g_z^{\rm G}(t_2-t_1;\Delta)dt_1dt_2+... \right],\label{SCyju}
\end{equation}
and then average $G_z^{\rm G}(t;\Delta,\nu)$ over $\rho(\Delta)$ in the similar manner as Eq~(\ref{GLconv}). The concept behind $\rho(\Delta)$ is that $\Delta$ can vary for each muon site (as illustrated in Fig.~3 of Ref.~\cite{Uemura:85}). 

However, the self-diffusion of muons and/or the motion of ions around them should be described as a stochastic process that effectively involves changes in muon sites. Therefore, when $\Delta$ is given by $\rho(\Delta)$ and ${\bm H}(t)$ undergoes the ``ergodic'' process in terms of fluctuations, the muon should perceive ${\bm H}(t)$ corresponding to a distribution with a different $\Delta=\Delta(t)$ for each fluctuation of ${\bm H}(t)$, so that Eq.~(\ref{SCyju}) should be rewritten as
\begin{equation}
G_z^{\rm G}(t;\Delta,\nu)=e^{-\nu t}\left[g_z^{\rm G}(t;\Delta_0)+\nu\int_0^t g_z^{\rm G}(t_1;\Delta_1)g_z^{\rm G}(t-t_1;\Delta_0)dt_1\right.\nonumber
\end{equation}
\vspace{-0.5cm}
\begin{equation}
\left.+\nu^2\int_0^t\int_0^{t_2} g_z^{\rm G}(t_1;\Delta_1)g_z^{\rm G}(t_2-t_1;\Delta_2)dt_1dt_2+... \right],\label{SCmodel}
\end{equation}
where $\Delta_i=\Delta(t_i)$ ($i=0$, 1, 2, ...).
To handle Eq.~(\ref{SCmodel}) statistically properly, it must be averaged over $\rho(\Delta)$ at this stage that reduces it to Eq.~(\ref{Gdyn}). This brings us back to the beginning without solving the problem. Conversely, assuming $\Delta$ remains constant during the evolution of the stochastic process implies the physically unlikely situation that the ${\bm H}(t)$ experienced by Mu does not reflect $\rho(\Delta)$ during the entire process. Thus it is concluded that the function obtained by Eq.~(\ref{SCyju}) does not provide a sound basis for the $\mu$SR data analysis.

The physically correct remedy to this problem is to introduce a cutoff in $n(H)$ at around $|H_{\rm max}|\propto 1/r_{\rm nn}^3$ determined by the minimal distance $r_{\rm nn}$ between the muon and the nn atoms [as schematically shown in Fig.~\ref{LKTfun}(b)], but at the cost of making the functional form more complicated, leading to difficulty to implement in practical analysis programs. As a next best solution, we approximate the overall relaxation function to the following phenomenological model in the motional narrowing regime:
\begin{equation}
 G_z^{\rm L}(t;\Lambda',\nu_{\rm f})\approx \exp\left[-\frac{2\Lambda^2}{\nu_{\rm f}}t\right], \label{ggz}
\end{equation}
where $\Lambda'=2\Lambda^2/\nu_{\rm f}$ is the linewidth incorporating the motional narrowing effect, and $\nu_{\rm f}$ is the fluctuation rate satisfying the condition $\nu_{\rm f}\gtrsim\Lambda$. Fig.~\ref{LKTfun}(c) and (d) show the dependence of $G_z^{\rm L}(t;\Lambda',\nu_{\rm f})$ on $\nu_{\rm f}$ and the corresponding change in $n(H)$. 
It is important to note that $G_z^{\rm L}(t;\Lambda,\nu)$ shown in Fig.~\ref{LKTfun}(a) does not overlap with $G_z^{\rm L}(t;\Lambda',\nu_{\rm f})$ except for $\nu=\nu_{\rm f}\approx\Lambda$. Moreover, $\Lambda'$ does not depend on LF. Therefore, the relation $\Lambda\approx\Delta$, which is the premise of Eq.~(\ref{Acf}), may be regarded as approximately restored in this approximation.

To verify that this is also true for $\nu\ll\Lambda$, let us examine the case of zero  field. Considering that the magnetic field felt by the muon is not zero but is given by the mean square of the internal magnetic field $\langle H^2\rangle\approx\Lambda^2/\gamma_\mu^2$ exerted from nuclear magnetic moments, the BPP relation Eq.~(\ref{bpp}) can be approximated as follows for the case of $\Lambda\approx\Delta$:
\begin{equation}
\Lambda\approx\frac{2\Delta^2\nu}{\nu^2+\Lambda^2}\approx\frac{2\Delta^2\nu}{\Lambda^2}\approx2\nu.
\end{equation}
The $\Lambda$ given here corresponds to the relaxation rate of the 1/3 component (whose relaxation rate is approximated by $\frac{2}{3}\nu$), and is consistent with the case of the dynamical Kubo-Toyabe function where $\Lambda$ is replaced with $\Delta$.

\renewcommand{\theequation}{B\arabic{equation}}
\subsection{Introduction of the Edwards-Anderson model}\label{EAsec}

In a disordered system such as amorphous glass, relaxation theories based on the single correlation time (e.g., that for NMR by Bloembergen, Purcell, and Pound~\cite{Bloembergen:48}) often fails to describe the time evolution of spin polarization, as has been demonstrated in the case of dilute spin glass \cite{Uemura:85}. Following the latter in which the coexistence of quasistatic and dynamical components was discussed, we introduce an Edwards-Anderson (EA) type autocorrelation function in place of  Eq.~(\ref{Acf}) \cite{Edwards:75,Edwards:76,Ito:24}:   
\begin{equation}
C_{\rm EA}(t)=\frac{\langle {\bm H}(0){\bm H}(t)\rangle}{\Lambda^2/\gamma_\mu^2} = \Phi e^{-\nu_{\rm f} t} +  (1-\Phi),
\label{AcfEA}
\end{equation}
where $\Phi$ is the EA parameter to represent the contribution of fluctuating component. Note that the effective linewidth becomes $\sqrt{\Phi}\Lambda$ for the fluctuating component and $\sqrt{1-\Phi}\Lambda$ for the quasistatic component, respectively. 

In applying the abovementioned model in the preliminary data analysis, it has been suggested that $\Phi$ is much smaller than unity at lower temperatures, so that the spin relaxation is dominated by the second term in Eq.~(\ref{AcfEA}).  On the other hand, $\Phi$ increases with increasing temperature above $T_{\rm d}\approx160$ K, suggesting that the first term in Eq.~(\ref{AcfEA}) becomes dominant with sharply increasing $\nu_{\rm f}$. The latter also suggests that Eq.~(\ref{ggz}) quickly approaches unity. In such situation, the relaxation function corresponding to Eq.~(\ref{AcfEA}) can be reasonably approximated by the following equation:
\begin{eqnarray}
G^{\rm EA}_z(t)& \approx& \zeta G_z^{\rm L}(t;\sqrt{\Phi}\Lambda',\nu_{\rm f})+(1-\zeta)G_z^{\rm L}(t;\sqrt{1-\Phi}\Lambda,\nu_{\rm f})\nonumber\\
& \approx& \zeta+(1-\zeta)G_z^{\rm L}(t;\Lambda_{\rm eff},\nu_{\rm eff}),
\label{gea}
\end{eqnarray}
where $\nu_{\rm eff}$ is the fluctuation rate of $\Lambda_{\rm eff}$ (with $\nu_{\rm eff}\lesssim\Lambda_{\rm eff}$), and $\zeta$ is the parameter to represent the relative contribution of the fast-fluctuating component.
\begin{figure}[t]
\includegraphics[width=0.95\linewidth,clip]{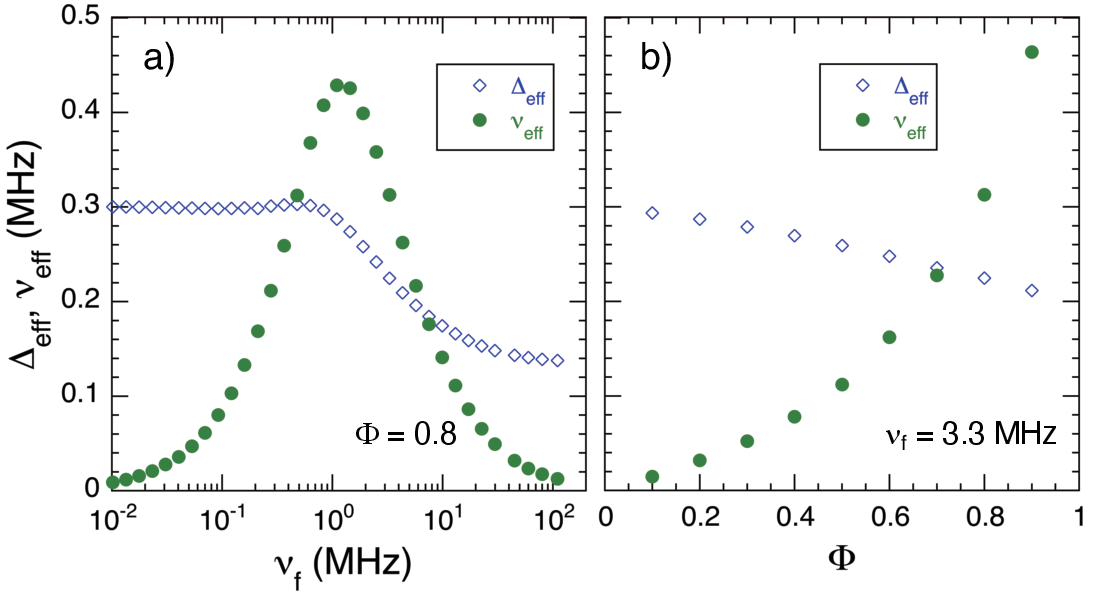}
	\caption{(a) Linewidth $\Delta_{\rm eff}$ and fluctuation frequency $\nu_{\rm eff}$ versus frequency $\nu_{\rm f}$ (in logarithmic scale) obtained by fitting the time spectra ($H=0$ and 1 mT) simulated by Gaussian $G_z^{\rm EA}(t)$ (with $\Lambda=0.3$ MHz, $\Phi=0.8$) using the conventional GKT function $G_z^{\rm G}(t;\Delta_{\rm eff},\nu_{\rm eff})$. (b) $\Phi$ dependence of $\Delta_{\rm eff}$ and $\nu_{\rm eff}$ at $\nu_{\rm f}=3.3$ MHz.}
	\label{ktfit}
\end{figure}

A key feature of this approximation is that it captures the variations of $G^{\rm EA}_z(t)$ associated with the crossover from one quasistatic limit for $\nu_{\rm f}\ll\Lambda$ (with $\Phi\approx0$, $\zeta\approx0$), 
$$G^{\rm EA}_z(t)\simeq G_z^{\rm L}(t;\Lambda,\nu_{\rm f})\approx\frac{1}{3}e^{-\nu_{\rm f} t}+\frac{2}{3}(1-\Lambda t)e^{-\Lambda t}$$ 
to another quasistatic limit for $\nu_{\rm f}\gg\Lambda$, 
$$G^{\rm EA}_z(t)\approx\zeta+(1-\zeta)\exp\left[-\frac{2\Lambda_{\rm eff}^2}{\nu_{\rm f}}t\right]g_z^{\rm L}(t;\sqrt{1-\Phi}\Lambda)$$
by the single LKT function $G_z^{\rm L}(t;\Lambda_{\rm eff},\nu_{\rm eff})$. 
While the exact $G^{\rm EA}(t)$ should be able to describe such variations in the lineshape with monotonically increasing $\nu_{\rm f}$,  $\nu_{\rm eff}$ in this approximation behaves such that it exhibits a maximum between the two limits. Given that $\nu_{\rm f}$ increases monotonically with temperature, $\nu_{\rm eff}$ increases almost proportionally to $\nu_{\rm f}$ while $\nu_{\rm f}\lesssim\Lambda$. However, further increase in $\nu_{\rm f}$ causes the crossover to the motional narrowing regime described by Eq.~(\ref{ggz}), leading to the increase in $\zeta$ and the decrease in relaxation rate of the second term. The latter is captured as an apparent decrease of $\nu_{\rm eff}$ in Eq.~(\ref{gea}), and the temperature dependence of $\nu_{\rm eff}$ exhibits a peak while the average fluctuation rate (dominated by $\nu_{\rm f}$) exhibits monotonous increase with temperature.

A similar behavior has been also observed in the analysis of $\mu$SR data in hybrid perovskites $A$PbI$_3$ using the conventional GKT function~\cite{Koda:22,Hiraishi:23,Papadopoulos:24}, and subsequently confirmed by the recent Monte Carlo simulations \cite{Kadono:26}. Specifically, using the GKT version of $G_z^{\rm EA}(t)$ generated numerically from Eq.~(\ref{AcfEA}) with $\Delta=0.3$ MHz, time spectra for various $\nu_{\rm f}$ were calculated for a given $\Phi$. Then these simulated spectra were analyzed via the $\chi^2$ fit using the approximation corresponding to Eq.~(\ref{gea}) (where $G_z^{\rm L}(t;\Lambda_{\rm eff},\nu_{\rm eff})$ was replaced with the conventional GKT function $G_z^{\rm G}(t;\Delta_{\rm eff},\nu_{\rm eff})$). As shown in Fig.~\ref{ktfit}, the deduced $\nu_{\rm eff}$ exhibits a peak at $\nu_{\rm f}\approx3\sqrt{\Phi}\Lambda$.  This occurs in parallel with a reduction of $\Delta$ for $\nu_{\rm f}>\Lambda$. On the other hand, in the case of thermally activated muon self-diffusion ($\Phi=1$), the accompanying fluctuation rate $\nu_{\rm f}$ increases monotonically and does not exhibit the aforementioned peak structure. Therefore, when the fluctuation rate deduced from the analysis using the conventional GKT function exhibits peaked temperature dependence, this can be regarded as experimental evidence that the motion of surrounding ions dominates as the cause of the fluctuation. The present case of LaH$_{3-2x}$H$_x$ is another example in which the local fields are described by the Lorentzian distribution.

Finally, we note that, for the GKT function, the precise lineshape of $G^{\rm EA}_z(t)$ can be numerically obtained via Monte Carlo simulations based on Eq.~(\ref{AcfEA}), and data analysis using musrfit with this numerical funciton is indeed feasible to directly deduce the relevant parameters \cite{Ito:24}. However, the LKT function inherently faces a fundamental difficulty upon taking the same approach: due to the non-physical properties of the Lorentzian distribution (see Appendix A), it cannot be used as the probability distribution function for the stochastic process in the simulation. Therefore, at present, only analysis using Eq.~(\ref{gea}) is feasible.

%
\end{document}